\documentclass{aa}

\usepackage{graphicx}
\usepackage{txfonts}
\usepackage{array}
\usepackage[percent]{overpic}
\usepackage{tcolorbox} % In the preamble
\usepackage{physics}
\usepackage{amssymb}
\usepackage{amsmath}
\usepackage{natbib}
\bibpunct{(}{)}{;}{a}{}{,} 
\usepackage{bm}
\usepackage{xcolor}
\usepackage[utf8]{inputenc}
\usepackage{tikz}

\newcolumntype{P}[1]{>{\centering\arraybackslash}p{#1}}
\newcolumntype{M}[1]{>{\centering\arraybackslash}m{#1}}
\usetikzlibrary{positioning,shapes.multipart,calc,shadows,arrows.meta}
\usetikzlibrary{decorations.pathreplacing}
\tikzset{
  basic/.style={draw, text centered},
  circ/.style={basic, circle, minimum size=2em, inner sep=1.5pt},
  rect/.style={basic, text width=1.5em, text height=1em, text depth=.5em},
  1 up 1 down/.style={basic, text width=1.5em, rectangle split, rectangle split horizontal=false, rectangle split parts=2},
}
\usepackage[colorlinks=true,     linkcolor=blue, citecolor=blue, filecolor=blue, urlcolor=blue]{hyperref}

\makeatletter
\let\linenumbers\relax
\makeatother

\begin{document}

%\title{OGLE-2023-BLG-0524: a free-floating planet candidate with legacy Hubble Space Telescope photometry}
%\title{Archival Hubble Space Telescope observations reveal a 25-year-old glimpse on the free-floating planet candidate OGLE-2023-BLG-0524}
\title{HST pre-imaging of a free-floating planet candidate microlensing event}
\titlerunning{OGLE-2023-BLG-0524}

\author{
    Mateusz Kapusta \inst{\ref{oauw}}
    \and Przemek Mr\'oz \inst{\ref{oauw}}
    \and Yoon-Hyun Ryu \inst{\ref{KASSI}}
    \and Andrzej Udalski \inst{\ref{oauw}}
    \and Szymon Koz\l{}owski \inst{\ref{oauw}}
    \and Sean Terry \vspace{0.2cm}\inst{\ref{Maryland},\ref{NASA}}\\ \vspace{0.2cm}
    Micha\l{} K. Szyma\'nski \inst{\ref{oauw}}
    \and Igor Soszy\'nski \inst{\ref{oauw}}
    \and Pawe\l{} Pietrukowicz \inst{\ref{oauw}}
    \and Rados\l{}aw Poleski \inst{\ref{oauw}}
    \and Jan Skowron \inst{\ref{oauw}}
    \and Krzysztof Ulaczyk \inst{\ref{Coventry},\ref{oauw}}
    \and Mariusz Gromadzki \inst{\ref{oauw}}
    \and Krzysztof Rybicki \inst{\ref{Weizmann},\ref{oauw}}
    \and Patryk Iwanek \inst{\ref{oauw}}
    \and Marcin Wrona \inst{\ref{Villanova},\ref{oauw}}
    \and Mateusz J. Mr\'oz \inst{\ref{oauw}} \\ {\it(The OGLE Collaboration)}\\ \vspace{0.2cm} 
     Michael D. Albrow\inst{\ref{canter}}
    \and Sun-Ju Chung \inst{\ref{KASSI}}
    \and Andrew Gould \inst{\ref{max_planck},\ref{astro_ohio}}
    \and Cheongho Han \inst{\ref{Cheongju}}
    \and Kyu-Ha Hwang \inst{\ref{KASSI}}
    \and Youn Kil Jung \inst{\ref{KASSI},\ref{UST}}
    \and In-Gu Shin \inst{\ref{Harvard}}
    \and Yossi Shvartzvald \inst{\ref{Weizmann}}
    \and Jennifer C. Yee \inst{\ref{Harvard}}
    \and Hongjing Yang\inst{\ref{westlake},\ref{Tsinghua}}
    \and Weicheng Zang \inst{\ref{Harvard}}
    \and Sang-Mok Cha \inst{\ref{KASSI},\ref{Yongin}}
    \and Dong-Jin Kim \inst{\ref{KASSI}}
    %\and Hyoun-Woo Kim
    \and Seung-Lee Kim\inst{\ref{KASSI}}
    \and Chung-Uk Lee\inst{\ref{KASSI}}
    \and Dong-Joo Lee\inst{\ref{KASSI}}
    \and Yongseok Lee \inst{\ref{KASSI}}
    \and Byeong-Gon Park\inst{\ref{KASSI}}
    \and Richard W. Pogge\inst{\ref{astro_ohio},\ref{cosmo_ohio}}
    \\ {\it(The KMTNet Collaboration)}
    }

\institute{Astronomical Observatory, University of Warsaw, Al. Ujazdowskie 4, 00-478 Warszawa, Poland \label{oauw} 
\and Korea Astronomy and Space Science Institute, Daejeon 34055, Republic of Korea \label{KASSI}
\and Department of Astronomy, University of Maryland, College Park, MD 20742, USA\label{Maryland}
\and Code 667, NASA Goddard Space Flight Center, Greenbelt, MD 20771, USA\label{NASA}
\and Department of Physics, University of Warwick, Coventry CV4 7 AL, UK\label{Coventry}
\and Department of Particle Physics and Astrophysics, Weizmann Institute of Science, Rehovot 76100, Israel \label{Weizmann}
\and Department of Astrophysics and Planetary Sciences, Villanova University, 800 Lancaster Ave., Villanova, PA 19085, USA \label{Villanova}
\and University of Canterbury, School of Physical and Chemical Sciences, Private Bag 4800, Christchurch 8020, New Zealand \label{canter}
\and Max-Planck-Institute for Astronomy, K\"onigstuhl 17, 69117 Heidelberg, Germany \label{max_planck}
\and Department of Astronomy, Ohio State University, 140 W. 18th Ave., Columbus, OH 43210, USA \label{astro_ohio}
\and Department of Physics, Chungbuk National University, Cheongju 28644, Republic of Korea\label{Cheongju}
\and National University of Science and Technology (UST), Daejeon 34113, Republic of Korea\label{UST}
\and Center for Astrophysics $|$ Harvard \& Smithsonian, 60 Garden St.,Cambridge, MA 02138, USA\label{Harvard}
\and School of Science, Westlake University, Hangzhou, Zhejiang 310030, China\label{westlake}
\and Department of Astronomy, Tsinghua University, Beijing 100084, China\label{Tsinghua}
\and School of Space Research, Kyung Hee University, Yongin, Kyeonggi 17104, Republic of Korea\label{Yongin}
\and Center for Cosmology and AstroParticle Physics, Ohio State University, 191 West Woodruff Ave., Columbus, OH 43210, USA\label{cosmo_ohio}
}

\date{Received XXX; accepted YYY}
\abstract{
    High-cadence microlensing observations uncovered a population of very short-timescale microlensing events, which are believed to be caused by the 
    population of free-floating planets (FFP) roaming the Milky Way. Unfortunately, the light curves of such events are indistinguishable from those caused by wide-orbit planets.
    To properly differentiate both cases, one needs high-resolution observations that would allow resolving a putative luminous companion to the lens long before or after the event.
    Usually, the baseline between the event and high-resolution 
    observations needs to be quite long ($\sim 10$ yr), hindering potential follow-up efforts. However, there is a chance to use archival data if they exist.
    Here, we present an analysis of the microlensing event OGLE-2023-BLG-0524, the site of which was captured in 1997 with the Hubble Space Telescope (HST).
    Hence, we achieve a record-breaking baseline length of $25$ years.
    A very short duration of the event ($t_\textrm{E}  = 0.346 \pm 0.008$ d) indicates an FFP as the explanation. We have not detected any potential companion to the lens with the HST data,
    which is consistent with the FFP origin of the event. Thanks to the available HST data, we are able to reject from $25\%$ to $48\%$ of potential stellar companions depending on the assumed population model.
    Based on the finite-source effects in the light curve we measure the angular Einstein radius value $\theta_\textrm{E} = 4.78\pm 0.23$ $\mu$as, suggesting a super-Earth in the Galactic disk or a sub-Saturn-mass planet in the Galactic bulge.
       %We search for other possible microlensing events, which can be captured on the same HST dataset, obtaining one close candidate.
    We show that the 
    archival high-resolution images should be available for several microlensing events, providing us with the unprecedented possibility of seeing the lensing system as it was many years before the event.
    
}

\maketitle
\section{Introduction}
Thousands of exoplanets have been identified since the first one was discovered at the end of the previous century \citep{mayor_1995}. The majority of them 
are bound to a host star. Nevertheless, many theoretical studies suggest the existence of planets that are gravitationally unattached to any star. They are also 
called free-floating planets, FFP for short. Their formation can 
occur in various ways (but not limited to): planet-planet scattering \citep*{rasio_dynamical_1996,lin_origin_1997,chatterjee_dynamical_2008}, interactions in 
stellar clusters \citep{spurzem_dynamics_2009} or post-main sequence evolution of planetary systems \citep{veras_great_2011}.

The introduction of high-cadence microlensing surveys allowed probing this elusive population of planets with masses ranging from that of Earth to that of Saturn
\citep{mroz_no_2017,gould_free-floating_2022,sumi_free-floating_2023}.
Contrary to stars and brown dwarfs, FFPs lead to very short-timescale ($t_\textrm{E} < 0.5$ d) microlensing events, where $t_\textrm{E}$ is the Einstein timescale.
Modern-day microlensing surveys such as Optical Gravitational Lensing Experiment (OGLE; \citealt{udalski_ogle-iv_2015}), Korea Microlensing Telescope Network (KMTNet, \citealt{kim_kmtnet_2016}) and
Microlensing Observations in Astrophysics (MOA, \citealt{bond_real-time_2001})
allowed measuring angular Einstein radii $\theta_\textrm{E}$ together with the relative lens-source proper motion $\mu_{\textrm{rel}}$ based on the 
finite-source effects for FFP-like events (for first such measurement see \citealt{mroz_2018}). Thanks to such observations we can safely rule out high proper motion as a cause for the short $t_\textrm{E}$ in such events,
further supporting the hypothesis of planetary-mass objects. A detailed review of FFP microlensing research together with the history of the discoveries is presented by \citet{mroz_exoplanet_2024}.

Even if light curves of short-timescale microlensing events seem to support the FFP hypothesis, it is hard to rule out the presence of a putative stellar companion at wide separation.
It is possible that the host star will never come close enough in the sky to the source star, leaving the light curve unaffected.
Fortunately, thanks to the relative lens-source proper motion, one can expect that after sufficient time the two will separate.
Then, there would be a possibility to investigate the lens's light with high-resolution imaging; either with the help of adaptive optics (AO) or with
space-based observatories. Such studies should reveal a putative host star to a planet if present. With a typical relative motion of $7$\,mas\,yr$^{-1}$ 
one needs to wait for several years before such inquiry would be possible. With the help of AO on a $30$-m-class telescopes one can expect around four times 
better resolution and hence four times shorter wait time \citep{gould_masada_2022} compared to the current generation of telescopes,
although right now we are inherently limited in our scope of tools. An example usage of AO techniques for FFP studies is
presented in \citet{mroz_free-floating_2024}, wherein several FFP microlensing events have been investigated.
Usually, such inquiry is based on the observations obtained long after the main event. There is however the possibility of using a pre-discovery data. 
There are no published application of this alternate
approach, although a similar strategy was proposed by \citet{kerins_magnifying_2023}, who suggest that the Euclid telescope can be used for pre-imagining of Roman Space Telescope fields in the Galactic bulge.
Such observations can effectively extend the baseline over which one can measure the proper motion of the stars and resolve source-lens pairs, which can add crucial information for the 
analysis of microlensing events. Moreover, such observations would allow for a much more rapid science return from the Roman Space Telescope mission as precursor 
imaging data could be obtained with Euclid.

In this publication, we present an investigation of the microlensing event OGLE-2023-BLG-0524. This very short-timescale microlensing event exhibits pronounced finite-source effects,
which can be used to measure its angular
Einstein radius. By pure coincidence, archival photometry from the Hubble Space Telescope (HST) is available and presents a high-resolution glimpse at the microlensing 
system as it was $25.55$ years before the event. This is a first time ever that such observations can be used to investigate the site of a microlensing event.
Due to unprecedented time span between the HST observation and the main microlensing event, one can expect that any potential putative host to the lens would be well separated on the images, 
allowing one to verify the hypothesis of FFP events origin. This paper is structured as follows. In the section 2, ground-based high-cadence photometry is presented, while section 3 is
devoted to the event modelling. Then, an overview of HST data is given in the section 4, followed by a detailed investigation that puts a limit on the putative host luminosity in section 5. 
Conclusions are presented in the section 6.

\section{Observations}
OGLE-2023-BLG-0524 was observed toward the Milky Way bulge on a star with equatorial coordinates $ (\alpha,\delta)_{\textrm{J2000}}=(18^\circ05'14.92'', -27^\circ59'14.0'')$,
which is located in the OGLE-IV field BLG511. The Galactic coordinates of the event are $(l,b) = (2.^\circ995,-3.^\circ248)$.
The maximum brightness was achieved on the 2023 May 22.
The event was alerted by OGLE Early Warning System \citep{udalski_optical_2003} on 2023 May 29.
The OGLE survey observed the field of interest every $60$ minutes during the night of the event with the $1.3$ m Warsaw Telescope located at Las Campanas Observatory, Chile. The telescope is equipped with 
a wide field-of-view $1.4$ deg$^2$ camera \citep{udalski_ogle-iv_2015}.
Only $I$-band observations are available for the night of the event.

This same object was observed with the Korea Microlensing Telescope Network (KMTNet, \citet{kim_kmtnet_2016}).
KMTNet uses three telescopes located at Cerro Tololo Interamerican Observatory (KMTC),
South African Astronomical Observatory (KMTS), Siding Springs Observatory (KMTA). Each facility hosts a $1.6$ m telescope with a $4$ deg$^2$ camera.
During the night of the event, the KMTC facility observed a field of interest roughly every $25$ minutes.
While almost all of the observations are conducted in the $I$ band, there are also a few observations in the $V$ band taken from KMTC.
Neither KMTNet AlertFinder \citep{AlertFinder} nor KMTNet Event Finder \citep{EventFinder}
found this event.

The photometric observations were reduced using custom pipelines by \citet{udalski_optical_2003} for OGLE and \citet{albrow_difference_2009} for KMTNet. Both pipelines 
are based on the Difference Image Analysis \citep*{tomaney_expanding_1996,alard_method_1998,wozniak_difference_2000}. Final KMTNet photometric observations were obtained with 
the KMTNet tender-love care pipeline \citep{KMT_tender}.

Only OGLE and KMTC were able to observe the event at the peak, KMTA captured a falling wing of the light-curve, while KMTS allowed to cover the baseline of the event
(For the light curve see Figure \ref{fig:FSFP_plot}).
The initial examination shows that brightening is extremely short. Despite being located in a region observed at high cadence, only seven and nine
observations have been made during the event by OGLE and KMTC, respectively. The event shows clear signs of the finite-source effects, 
which suppress the maximum brightness in the peak to only $\sim 2$\,mag.

\section{Light curve modelling}
\begin{figure}
    \centering
    \includegraphics[width = 0.5\textwidth]{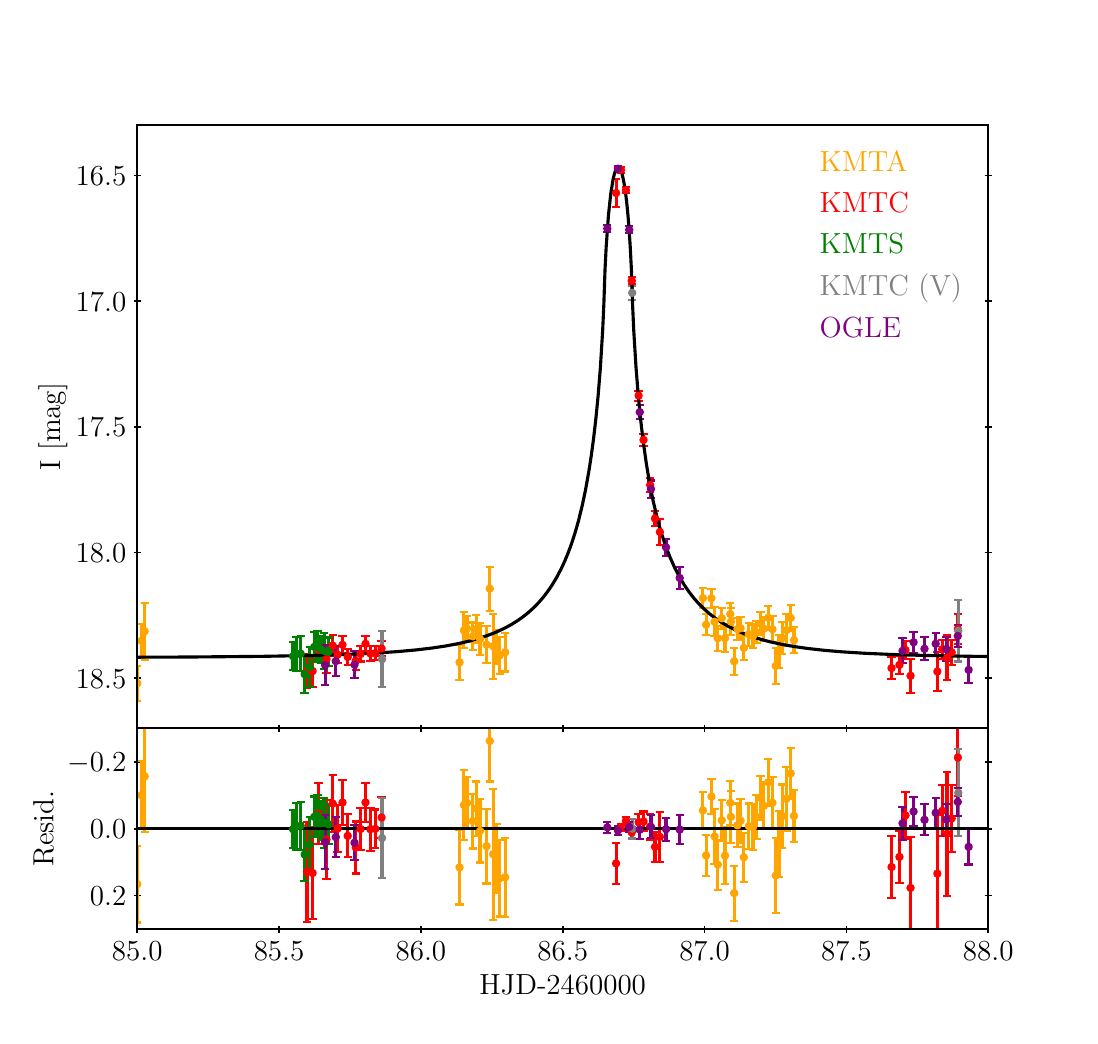}
    \caption{The light curve of OGLE-2023-BLG-0524 together with the best-fit FSPL model (black solid line). All observations are scaled to the OGLE system.}
    \label{fig:FSFP_plot}
\end{figure}

%In the fitting procedure all available datasets were used (we are limiting ourselves to observations collected in 2022 and 2023).
We used all available OGLE and KMT datasets in the fitting procedure, limiting ourselves only to the data collected in 2022 and 2023.
The standard point-source point-lens model (PSPL) is defined using the time of maximal brightening $t_0$, the Einstein timescale $t_\textrm{E} = \theta_{\textrm{E}}/\mu_{\textrm{rel}}$, and 
 the impact parameter $u_0$ (expressed in terms of the Einstein radius). Finite-source effects are parametrized with $\rho = \theta_*/\theta_{\textrm{E}}$,
where $\theta_*$ is the angular radius of the source star.
In order to suppress any correlations between variables, parameters $t_* = t_\textrm{E} \rho$ and $b_0 = u_0/\rho$ have been defined. Both variables, 
together with $t_0$, $\rho$ were used to fully define the light-curve model.
During the fitting, we decided to calculate the best-fitting source flux $F_s$ and blending flux $F_b$ for a set of the parameters to compute the $\chi^2$ value.
As there is little to no information about the limb darkening
coefficient at this stage, it was decided to use $\Gamma = 0.45$, which should be valid for a K-type star.
The Python-based library \texttt{emcee} \citep{foreman-mackey_emcee_2013} has been used to sample from the posterior and estimate 
uncertainties associated with the parameters.
Inferred parameters together with derived ones are listed in Table \ref{tab:par_1}. The light curve of the event together with theoretical predictions are presented in 
Figure \ref{fig:FSFP_plot}.
The estimation based on the microlensing models allows us to measure the calibrated $I$ magnitude of the source star, which is equal to $I = 19.72^{+0.19}_{-0.24}$ mag.

The source-to-baseline flux ratio $f_{s} = F_s/(F_s + F_b)$ suggest significant 
blending for all datasets. This important parameter is equal to $f_{s,\textrm{KMTC}} = 0.63_{-0.10}^{+0.15}$ and 
$f_{s,\textrm{OGLE}} = 0.30_{-0.05}^{+0.07}$ for KMTC and OGLE datasets respectively.
Significant blending, in principle, may be attributed to the luminosity of the lens itself or companion to the source.
This rule, however, is not absolute, as the blending flux could also originate from nearby field stars.

\begin{table}
\begin{center}
    \def\arraystretch{1.4}
    \caption{Initial microlensing parameters of the event grouped by sampling parameters (upper part) and derived parameters including blending parameters for datasets (lower part).}
    \label{tab:par_1}
    \begin{tabular}{ l|c }
        \hline 
        Parameter & Value\\
        \hline
        $t_0$ (HJD$-2460000$) & $86.69579\pm 0.00072$ \\
        $t_*$ (d) & $0.0504^{+0.0012}_{-0.0010}$ \\
        $b_0$ & $0.194^{+0.126}_{-0.112}$\\
        $\rho$ & $0.119_{-0.018}^{+0.024}$\\
        \hline 
        \hline
        $u_0$ & $0.025^{+0.021}_{-0.015}$\\
        $t_\textrm{E}$ (d) & $0.422_{-0.065}^{+0.067}$\\
        $f_{s,\textrm{KMTA}}$ & $0.47_{-0.12}^{+0.19}$\\
        $f_{s,\textrm{KMTC}}$ & $0.63_{-0.10}^{+0.15}$\\
        $f_{s,\textrm{KMTS}}$ & $0.31_{-0.12}^{+0.24}$\\
        $f_{s,\textrm{KMTC-V}}$ & $0.32_{-0.05}^{+0.08}$\\
        $f_{s,\textrm{OGLE}}$ & $0.30_{-0.05}^{+0.07}$\\
        \hline
    \end{tabular}
\end{center}

\end{table}

\section{Hubble Space Telescope data}
As the region of the event is very close to the center of the Galactic bulge, it was observed by several microlensing surveys during the last $30$ years. In 1996, the  MACHO collaboration 
observed an event that later got designated as  MACHO-96-BLG-14 \citep{alcock_macho_2000}. Following the discovery of the event, it was observed using 
the Wide Field and Planetary Camera 2 (WFPC2) with the HST \citep{bennett_prop}. In total, five images were taken on 1997 Nov 1: three in the F555W filter and two in the F814W filter (equivalents 
of Johnson $V$ and Cousins $I$ filters, respectively). Each exposure lasted only $40$ s.
The WFPC2 camera consisted of four CCD detectors (numbered from 1 to 4), each with $800 \times 800$ pixels. 
The first detector (Planetary Camera, PC for short) was two times smaller than the other three, with the same number of pixels, making its resolution effectively two times  
higher than other detectors. In the follow-up observations of MACHO-96-BLG-14, the Planetary Camera was used to observe the main event, which was only a few arcminutes away
from the star that would become 
the source for OGLE-2023-BLG-0524 more than $25$ years later. Fortunately, the HST captured the site of OGLE-2023-BLG-0524 on the second detector in the WFPC2 camera.
The effective size of the pixel in the second detector is around $0''.0996$, allowing for a sub-arcsecond resolution of the event.
Approximately 9333 days ($25.55$ yr) passed between the HST observations and the microlensing maximum.
%Between the HST observations and the microlensing maximum approximately 9333 days ($25.55$ yr) passed.

To mitigate the chance for potential mistake in the source identification, on 2025 April 26, we obtained additional two orbits of HST observations from the program GO-17834 \citep{HST_new},
using the WFC3/UVIS camera. This was 1.93 years after the peak of the microlensing event.
We obtained 16 x 69 sec. dithered exposures with the F814W filter and 16 x 70 sec. dithered exposures with the F606W filter using the UVIS2-C1K1C-SUB aperture.
We use this sub-array to minimize the degradation effect of charge transfer efficiency (CTE) in these HST detectors.
The analysis is performed using the PSF-fitting routine hst1pass \citep{hst_pass}, which generates a distortion-free point source catalog of all detected stars,
including the OGLE-2023-BLG-0524 source star.

\subsection{OGLE-HST astrometric transformation}
We found that the equatorial coordinates provided with the WFPC2 image deviated by around $3''$ between the OGLE and HST positions.
Hence, we decided to match the OGLE and HST images without the help of supplemented coordinates.
In order to locate the microlensing star, we established a transformation between the coordinates on the OGLE reference image and those on the HST image.
We used \texttt{DOLPHOT} \citep{dolphin_dolphot_2016} to measure the position and brightness of stars on the HST image.
We decided to start with a basic linear transformation realized in the form of rotation and shift. OGLE uses pixels 
with size $0''.26$, which are roughly $2.5$ times bigger than those used in WFPC2. Hence, an additional scale parameter was added that scales relative positions on 
the OGLE image to those on the HST image, resulting in a total of four parameters to fit.
Then, the first image in the F814W band was selected as a grid reference. We decided to select only stars brighter than $18.5$ mag to calculate the coefficients of the transformation.
Each OGLE star brighter than this limit was cross-matched with the Gaia DR3 catalog \citep{gaiacollaboration_gaia_2023} allowing us to obtain its proper motion.
%Those objects were ''returned'', where they should be located at the time when the HST image was taken.
We then calculated the proper-motion-corrected positions of the stars for the epoch of the HST observations.
Subsequently, the coordinates of each star were transformed to a WFPC2 base with an additional correction for a WFPC2 camera distortion \citep{casetti-dinescu_comprehensive_2021}.
With the established transformation, the mean distance between the nearest stars on the HST image and the transformed OGLE object was calculated and minimized using the Nelder-Mead algorithm.
Finally, the transformation was established with the mean root square error (RMS)
of $0.8$ HST pixel. To assess the limitations of the four-parameter linear model, we tested more complex transformations,
including a six-parameter linear model and a quadratic transformation. All transformations performed similarly, yielding comparable RMS errors and final object positions.

The position of the baseline star of OGLE-2023-BLG-0524 is $(X_\textrm{O},Y_\textrm{O}) = (723.23,351.08)$ on the OGLE reference image. 
We measured the accurate position of the source star using subtracted images 
in the peak obtaining $(X_\textrm{O}',Y_\textrm{O}') = (720.950 \pm 0.055,351.565 \pm 0.093)$. After the transformation to the HST frame, we obtained coordinates $(X_\textrm{H},Y_\textrm{H}) = (578.81, 247.59)$.

We do not report any problems with 
positional coordinates supplied with the WFC3 data. 
We measure the position of the event on the WFC3 dataset with supplied positional information.
Coordinates of the source star on the detector are $(X,Y) = (513.49\pm 0.02,620.18 \pm 0.02)$.

\begin{figure*}

    \begin{overpic}[width=\textwidth]{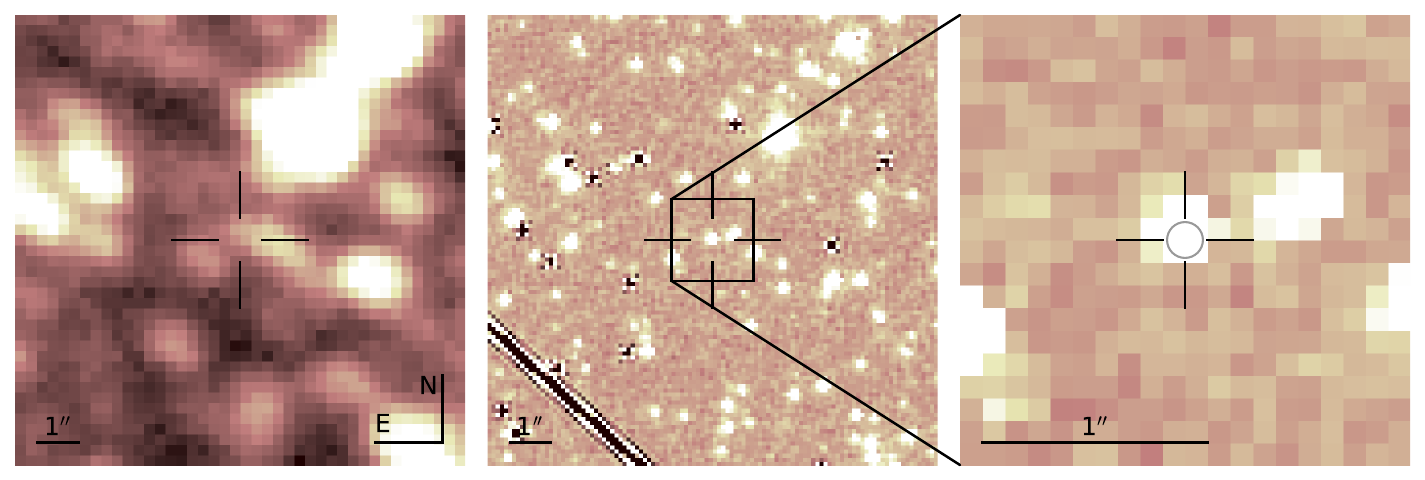}
        \put(70,30){
        \begin{tcolorbox}[width=5cm,colback=blue!10,colframe=blue!80!black,boxrule=0.3mm,arc=2mm,auto outer arc, left=2pt, right=2pt, top=2pt, bottom=2pt ]
            { \color{black}WFPC2 (1997) - F814W}
        \end{tcolorbox}}
    \end{overpic}
    \begin{overpic}[width=\textwidth]{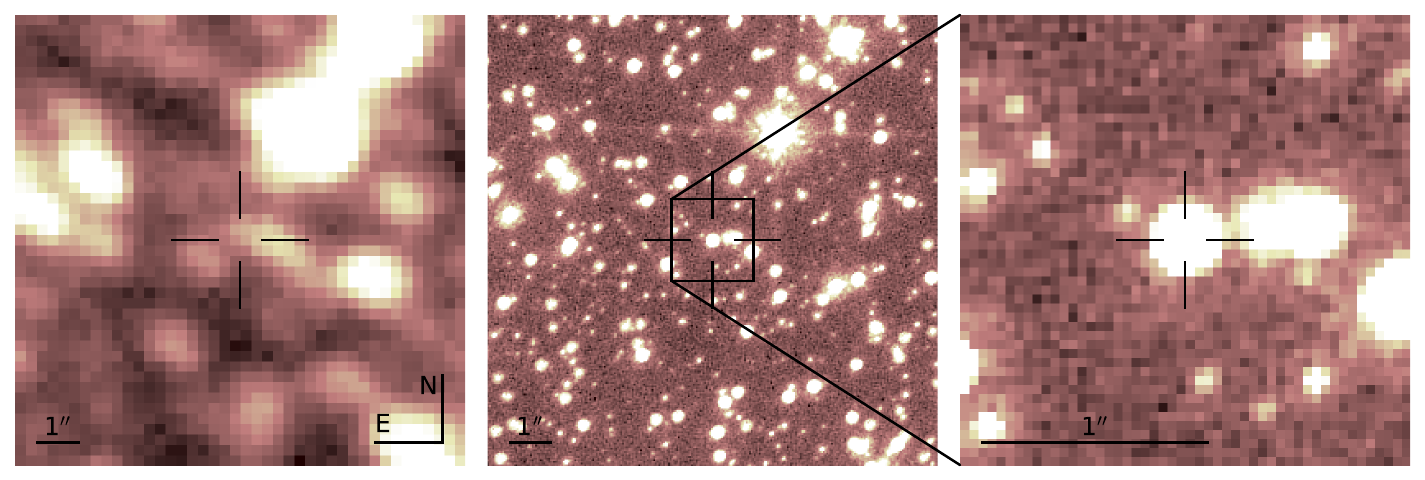}
        \put(70,30){
        \begin{tcolorbox}[width=5cm,colback=blue!10,colframe=blue!80!black,boxrule=0.3mm,arc=2mm,auto outer arc, left=2pt, right=2pt, top=2pt, bottom=2pt ]
            { \color{black} WFC3-UVIS (2025) - F814W}
        \end{tcolorbox}}
    \end{overpic}
    \caption{A finding chart of OGLE-2023-BLG-0524 event viewed by OGLE (left panel) and HST (middle and right panel).
    The two upper-right panels present archival WFPC2 data, while the two lower-panels contain modern WFPC3 dataset.
    A circle represents the RMS error of the linear transformation (only in the upper-right panel). The Hubble images were rotated and rebinned to 
    align the images with the North-East direction.
    }
    \label{fig:chart}
\end{figure*}

A detailed view of the vicinity of the event (on both HST datasets) is presented in Figure \ref{fig:chart}. As we can clearly see, the field is very crowded.
This provides us with an easy explanation for the observed blending in the microlensing model.
The source star in the OGLE reference image is composed of at least three HST counterparts separated by around $500$ mas.
%Hence, the chance that the source star was misidentified is minimal.

\subsection{HST view on the source star}
We used \texttt{DOLPHOT} \citep{dolphin_dolphot_2016} to measure the brightness of the source star on the WFPC2 image. The brightness in F814W and F555W filters is equal to $\textrm{F814W}=19.484 \pm 0.023$ mag and 
$\textrm{F555W} = 21.112 \pm 0.029$ mag, respectively. Using transformations from WFPC2 to UBVRI presented in \citet{holtzman_photometric_1995} we obtained
$V-I = 1.670 \pm 0.038$ mag, $I = 19.442 \pm 0.023$ mag and $V = 21.112 \pm 0.029$ mag. We decided against establishing our own transformation between OGLE and WFPC2 magnitudes due 
to the systematic effects of blending, which may cause problems for the potential transformation. Despite nearly $0.3$ mag difference between the HST value and one
obtained from the microlensing source estimate, both values are still consistent within $1.2\sigma$.
There is however a possibility that this difference in magnitudes is real and an additional $20.94_{-0.41}^{+0.85}$ mag star is contributing to the total luminosity of the HST star.
This additional light may come from the companion to the source or the luminous lens

To obtain the magnitude of the source star in WFPC3 observations, we cross-matched WFC3 and WFPC2 images to calibrate the brightness.
We fitted the linear model to find the magnitude in the WFPC2 F814W filter based on the WFC3 F606W and F814W magnitudes. We found $F814W' = 19.493\pm0.014$, which is 
in good agreement with the WFPC2 value.

\begin{figure}
    \includegraphics[width = 0.5\textwidth]{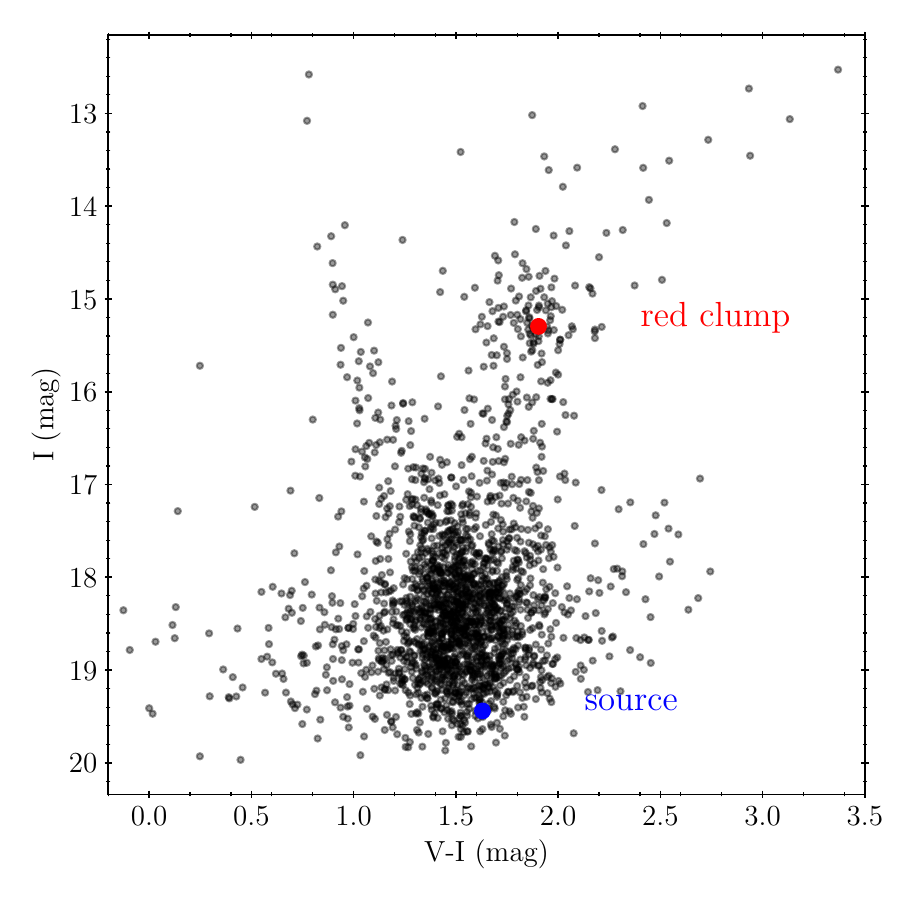}
    \caption{Color-magnitude diagram showing stars in the vicinity of the event ($2'\times 2'$).}
    \label{fig:cmd}
\end{figure}

Then we proceeded with the standard method presented in \citet{yoo_ogle-2003-blg-262_2004}, which allowed us to de-redden our photometry and calculate 
the angular radius of the source star.
The red clump centroid in the vicinity of the event is $((V-I)_{\textrm{RC}},I_{\textrm{RC}}) = (1.904 \pm 0.015,15.296 \pm 0.025)$ (presented in Figure \ref{fig:cmd}),
while de-reddened values are equal to $((V-I)_{\textrm{RC,0}},I_{\textrm{RC,0}}) = (1.06,14.350)$ \citep{nataf_reddening_2013,Bensby_2013}.
Assuming that the reddening toward the source is the same as that toward red clump stars, the de-reddened color 
of the source is equal $(V-I)_0 = 0.826 \pm 0.041$, which can be translated to around $5300 \pm 100 $ K and $(V-K)_0 = 1.842 \pm 0.092$ using the relation presented in \citet{ramirez_effective_2005}.
In order to find the final microlensing parameters, the new limb darkening value was calculated with the help of \citet{claret_gravity_2011} and is equal
$\Gamma = 0.44$ for observations in the $I$ band and $\Gamma = 0.62$ for observations in the $V$ band (for a main sequence star). We employed the color-surface brightness relationship presented in \citet{adams_predicting_2018},
allowing us to compute the angular diameter for a source.
Then, new microlensing models were created: with and without a prior on the OGLE source magnitude. We assumed a Gaussian prior on $I_s \sim \mathcal{N}(19.442,0.023)$.
The comparison between parameters of both samplings together with the expected lens-source separation $25.55$ years before the peak is presented in Table \ref{tab:par_2}.

\begin{table*}
    \def\arraystretch{1.4}
    \caption{Microlensing parameters computed with and without a prior on the source's star magnitude.}
    \label{tab:par_2}
    \begin{center}
        \begin{tabular}{l c c}
            \hline
            \hline
            \par & w/t prior & w/ prior \\
            \hline
            Microlensing model & \par &\par \\
            \hline
            $t_0$ (HJD$-246000$) & $86.69562 \pm 0.00062$ & $86.69560 \pm 0.00063$ \\
            $t_*$ (d) & $0.0504^{+0.0011}_{-0.0010}$ & $0.05162 \pm 0.00071$ \\
            $b_0$ & $0.20 \pm 0.13$ & $0.318^{+0.044}_{-0.055}$ \\
            $\rho$ & $0.119^{+0.024}_{-0.018}$ & $0.1492\pm 0.0034$ \\
            $t_\textrm{E}$ (d) & $0.429 \pm 0.068$ & $0.3460 \pm 0.0083$ \\
            $u_0$ & $0.023_{-0.016}^{+0.021}$ & $0.0474_{-0.0079}^{+0.0062}$\\
            $\chi^2/\textrm{d.o.f.}$ & $6660.86/6359$ & $6661.65/6359$ \\ % 6373 points
            \hline
            Source star parameters & \par &\par \\
            \hline
            $I_S$ & $19.732_{-0.235}^{+0.193}$ & $19.444 \pm 0.023$\\
            %$(V-I)_S$ & $1.10_{-0.02}^{+0.03}$ & $1.11_{-0.02}^{+0.03}$\\ 
            $\theta_*$ ($\mu$as) & $0.626 ^{+0.075}_{-0.059}$ & $0.713\pm 0.033$ \\
            \hline
            Physical parameters & \par &\par \\
            \hline
            $\theta_\textrm{E}$ ($\mu$as) & $5.25^{+0.47}_{-0.44}$ & $4.78\pm 0.23$\\
            $\mu_{\textrm{rel}}$ (mas\,yr$^{-1}$) & $4.54^{+0.46}_{-0.39}$ & $5.04 \pm 0.25$\\
            separation (mas) & $117^{+13}_{-12}$ & $129 \pm 8$ \\   
            \hline
        \end{tabular}
    \end{center}

\end{table*}

\section{Detectability Simulations}
As it was derived in the previous sections, one can expect that the source and the lens should be separated by around $129$ mas, which can be translated to $1.3$
pixels on the HST image taken $25.55$ years earlier. No such object was found by \texttt{DOLPHOT}, supporting the FFP hypothesis.
Nevertheless, the FFP hypothesis may not be correct, as the faint star may be undetectable 
on the HST image due to the presence of a nearby field star. In order to investigate limits on a possible companion we decided to simulate its detectability in the HST images.
The nearest object, located just five pixels away, is actually composed of two stars separated by only one pixel.
This stellar asterism can complicate the detection problem, as one expects that detectability will not be independent of the source-lens angle. 

This section is structured as follows.
In section 5.1, a theoretical description of the detectability simulations is presented. Subsequently, three subsections are devoted to particular detection cases.
In section 5.2, we use tools described in the first subsection to model the actual HST image. Then, in sections 5.3 and 5.4 
we test the detectability limits of putative host stars.

\subsection{Theoretical introduction}
Let us denote the total photons counted in pixel $(X,Y)$ as $N(X,Y)$, while the mean number of theoretically predicted counts is designated $F(X,Y)$.
The simulation begins with the initialization of the $12 \times 12$ grid (which covers roughly $1''\times 1''$). Then, $N(X,Y)$ is
sampled from a Poisson distribution with rate $F(X,Y)$. At the end, we add additional Gaussian noise $\sigma$ to each pixel, which simulates the readout noise.
For a given separation between stars $s$ and their magnitude difference $m$ we fit two models: the single-star model and the double-star model. We perform fitting with the
Nelder-Mead algorithm using $\chi^2$ as a loss function
\begin{equation}
    \chi^2 = \sum_{X,Y} \frac{(F(X,Y)-N(X,Y))^2}{N(X,Y) + \sigma^2}.
\end{equation}

It was decided to use the \texttt{TinyTim} \citep{krist_20_2011} PSF library to simulate PSFs for all models. 
This particular approach is better than \texttt{DOLPHOT}, as TinyTim numerically computes the PSF rather than relying on simple analytical approximations.
As the size of the PSF is determined by the position on the detector, we only need to fit the position of the star. After using maximally sub-sampled PSF,
we perform the linear interpolation with re-binning, followed by a convolution with a charge diffusion kernel, which is presented in the TinyTim manual
\footnote{https://www.stsci.edu/files/live/sites/www/files/home/hst/\\instrumentation/focus-and-pointing/documentation/\_documents/\\tinytim.pdf}.
In the single-star model, we include only one star with a constant sky value. We have four parameters in total: position ($X_1$,$Y_1$), total luminosity of star 
$F_1$ and background $B$. In the double-star model, we add an additional star with three more parameters: $(X_2,Y_2)$ and $F_2$.
In general,one can write that. 
\begin{equation}
    F(X,Y) = \sum_i F_i \cdot \textrm{PSF}(X-X_i,Y-Y_i) + B,
\end{equation}
where we have summation over the number of stars and where PSF denotes the point spread function obtained from \texttt{TinyTim}. 

\subsection{Modelling the HST image}
As we noted in previous sections, the source star is much fainter in the F555W filter than in the F814W one. This is reflected by the higher source-to-noise ratio (SNR) in the F814W filter,
which is around $34$ compared to $22$ in the case of F555W.
In order to maximize the chances of detection, we decided to work on images in the F814W filter only.
Observations have been made in "$15$" gain mode, indicating a gain value of around $\sim14.5$ and readout noise $\sim7.84$
(taken from WFPC2 manual\footnote{https://www.stsci.edu/files/live/sites/www/files/home/hst/ \\ documentation/\_documents/wfpc2/wfpc2\_ihb\_cycle17.pdf}).
A $12 \times 12$ cutout was created from the image, which
contains the source star and nearest neighbor clump of stars.

\begin{figure}
    \begin{center}
            \includegraphics[width = 0.5\textwidth]{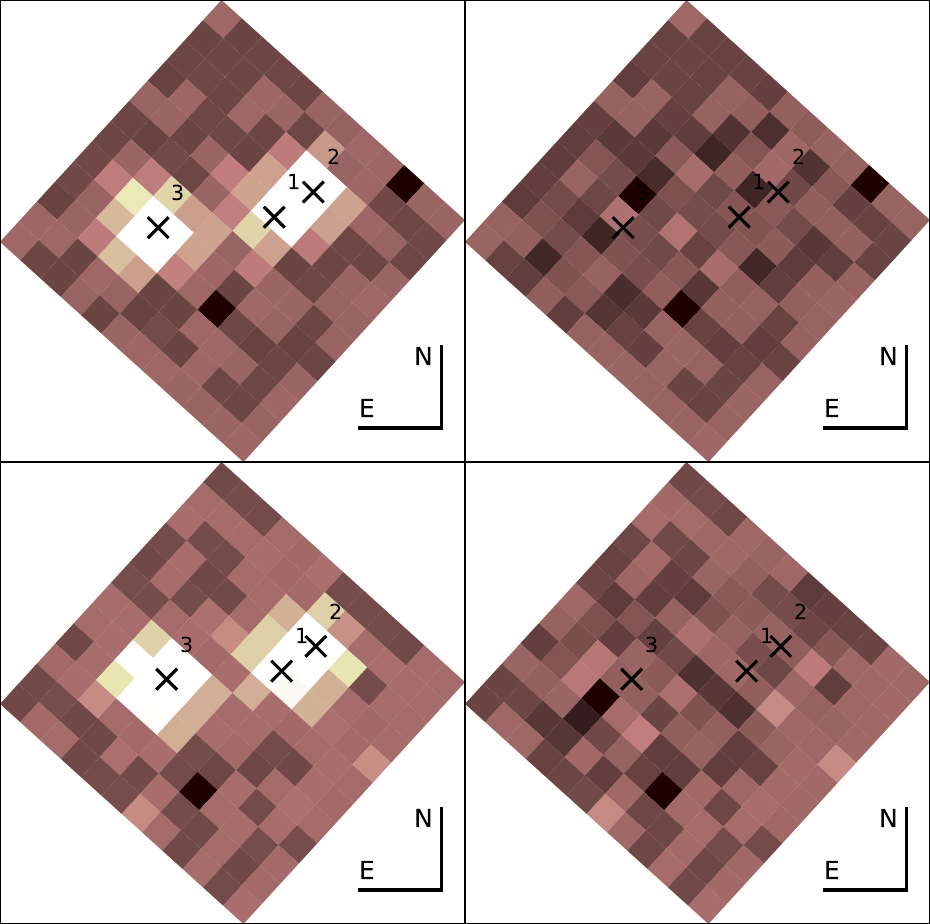}
    \caption{F814W residuals together with images used for the fitting. Each star is marked with a number according to its luminosity (from darkest to brightest).
    Microlensing source is the brightest and is denoted with the number "3". Images are rotated to align with the North-East direction.}\label{fig:residuals}
    \end{center}
\end{figure}

We know that the \texttt{DOLPHOT} was able to detect only three stars in the cutout. Hence, it was decided to fit 
four and three-star TinyTim PSF models. With the three-star model we should obtain a similar fitting to the  \texttt{DOLPHOT} model.
With the four-star model we can search for the putative companion to the lens. 
The three-star model achieved $\chi^2 = 140.4$ and $\chi^2 = 150.4$ for $132$ and $133$ degrees of freedom on the first and second image, respectively
(two and one pixels on the first and second image respectively were reported as bad and hence they were not used in the analysis).
The four-star model is not able to improve the score 
beyond $\Delta \chi^2 = 2$, so no significant detection is reported. Images used in the inquiry together with residuals from the three-star model are presented in Figure 
\ref{fig:residuals}. The source star's brightness was measured and is equal $F_s = 2125.6$ e$^{-}$, while the background is equal to around $B = 20.6$e$^{-}$.
Those two values are then used to perform simulations.

%\subsection{Is the HST source star blended?}
%First of all, we need to ask ourselves whether there is a chance that the source star in the WFPC2 image is blended. This is an important 
%question, as the posterior of the microlensing model predicts a dimmer source star than the observed object, albeit at low significance.
%In order to resolve the issue we selected a subsample from the inference chain,
%with a flat prior on the source's brightness.
%In general, fainters stars greatly outnumber those brighter in considered case so the assumption about the prior is not realistic.
%Nevertheless, this type of analysis aims to give a rough estimate whether we can spot this type of blending star.
%If the blending light originates from the putative companion to the lens, we can simulate, whether we would be able to detect such a star. 
%We used the sub-sampled chain to simulate the potential population of blending stars.
%To simplify the situation, simulations only included two stars, with total luminosity equal to the one measured with \texttt{DOLPHOT}. 
%The relative angle between the stars was randomized.
%It turns out that only $82\%$ of such objects can be detected in a simulation. Hence, there is no certainty that use of the HST source
%star prior in the microlensing model is correct. This particular value can be also subjected to some fluctuations associated with the 
%choice of prior.
%In the final analysis we decided to use a HST prior on the source's magnitude, as the significance of blending light is low.

\subsection{Detecting putative host star}
We can use the HST image to perform injection-and-recovery simulations and find out whether the putative host star is detectable.
Such simulations are composed of two steps: injection of the putative star into a simulated image and fitting a model to recover the injected object.
First, we need to create artificial images containing two stars (source and putative companion to the lens), each with dimensions $12 \times 12$.
Then, we fit a single-star model and a double-star model, and compare the goodness of fit for both models.

To create an artificial population of host stars, we 
need to specify the relative host-source separation and the luminosity distribution of the stars. 
We used distribution of separations from our Bayesian microlensing model
(second column in Table \ref{tab:par_2}).
We assumed that the total luminosity of stars is constant
and equal to $I = 19.484$.
While the total brightness of stars was constant, the relative magnitude difference was sampled 
from the uniform distribution $\Delta m \sim \mathcal{U}(0,5)$.

\begin{figure}
    \begin{center}
        \includegraphics[width = 0.5\textwidth]{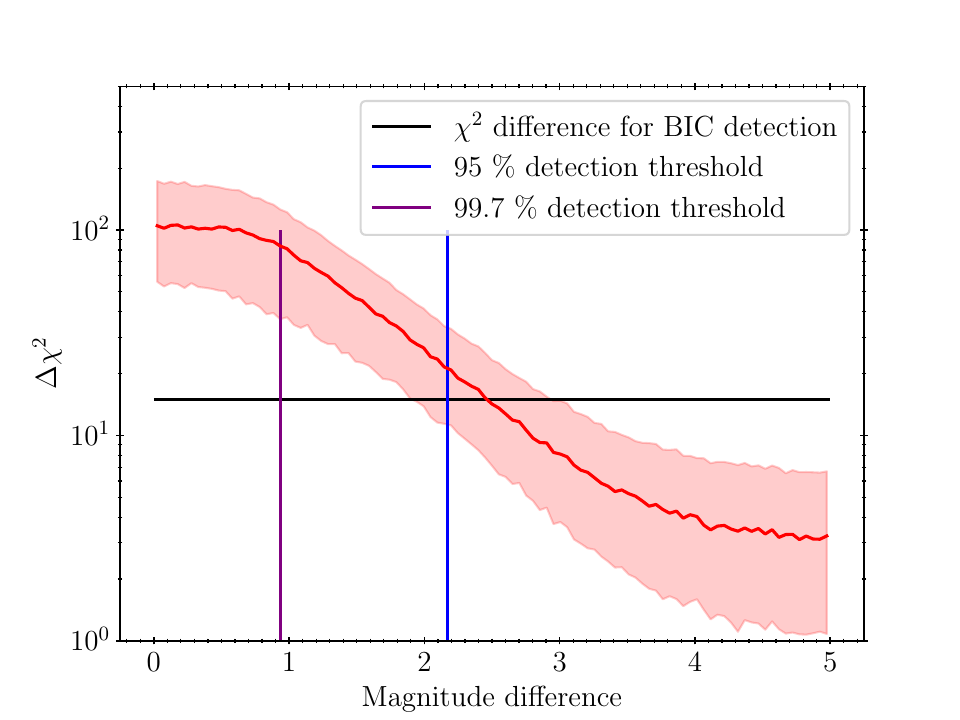}
        \caption{Simulated $\Delta \chi^2$ differences (with  $1\sigma$ interval) plotted against the magnitude difference between both stars. Vertical lines indicate 
        $2\sigma$ and $3\sigma$ detection thresholds.}\label{fig:const_r}
    \end{center}
\end{figure}

In total $2\cdot10^5$ simulations have been performed. The relative angle between the stars was randomized.
We decided to use a BIC score to determine which model is preferred. The BIC score is defined by the formula 
\begin{equation}
    \textrm{BIC} = \chi^2 + k \ln{n},
\end{equation}
where $k$ is the number of parameters while $n$ denotes the size of our image in pixels (here $144$, we use a $12 \times 12$ grid).
If the BIC score for a two-star model ($\textrm{BIC}_2$) is lower than for the one-star model ($\textrm{BIC}_1$) we would prefer the two-star model.
Usually, one requires a difference of about a dozen in the BIC scores to decide which model is preferred.
However, we decided not to raise the bar for detection, so the lower BIC score for the double-star model would suffice.
If the double-star model is better, we can write that 
\begin{equation}
    \textrm{BIC}_1 - \textrm{BIC}_2 = \Delta \textrm{BIC} = \Delta \chi^2 - 3\ln{144} > 0,
\end{equation}
as the double-star model has three more parameters than the single-star model.
%A subscript on a value indicates the number of stars in the model used to compute it.
We defined the difference in $\chi^2$ scores
as $\Delta \chi^2 = \chi^2_1 - \chi^2_2$, where $\chi^2_1$ and $\chi^2_2$ denote $\chi^2$ scores for the single and double model, respectively. 
Hence, we require $\Delta \chi^2 > 3\ln{144} \approx 14.9$ difference to report a detection.
Simulated $\Delta \chi^2$ differences between a two-star model and a one-star model are presented in Figure \ref{fig:const_r}. Aditionally, a $1\sigma $ interval has been over-plotted.
This intrinsic uncertainty is associated with a different initialization of the grid and different separations.
%Let us define a magnitude difference $m_t$ that $95\%$ of host stars
%with a magnitude difference, smaller than $m_t$ are detected.

Usually, one would like to assess whether stars with luminosity above some threshold are detectable or not.
This division becomes blurry due to the very low signal-to-noise ratio of the source-star, which prompted us to establish another form of 
detectability criterion.
Let $\Delta m_t$ be a magnitude difference that $95\%$ of putative host-source pairs with relative brightness $\Delta m < \Delta m_t$ are detected.
Then, the $\Delta m_t$ value can be used to calculate the corresponding magnitude of the host star $m_t$ as we know the total luminosity of stars. 
Even if the value $\Delta m_t$ is clearly defined, it depends on the choice of our prior on the luminosity and separation. 
While the separations are modeled according to the posterior distribution of separations from the microlensing model, the 
distribution of the relative magnitudes is described in a somehow artificial manner. Nevertheless, as we were mainly concerned with a simple estimation, 
we settled on the simple uniform distribution as noted in the previous paragraph.

\begin{figure}
    \includegraphics[width = 0.5\textwidth]{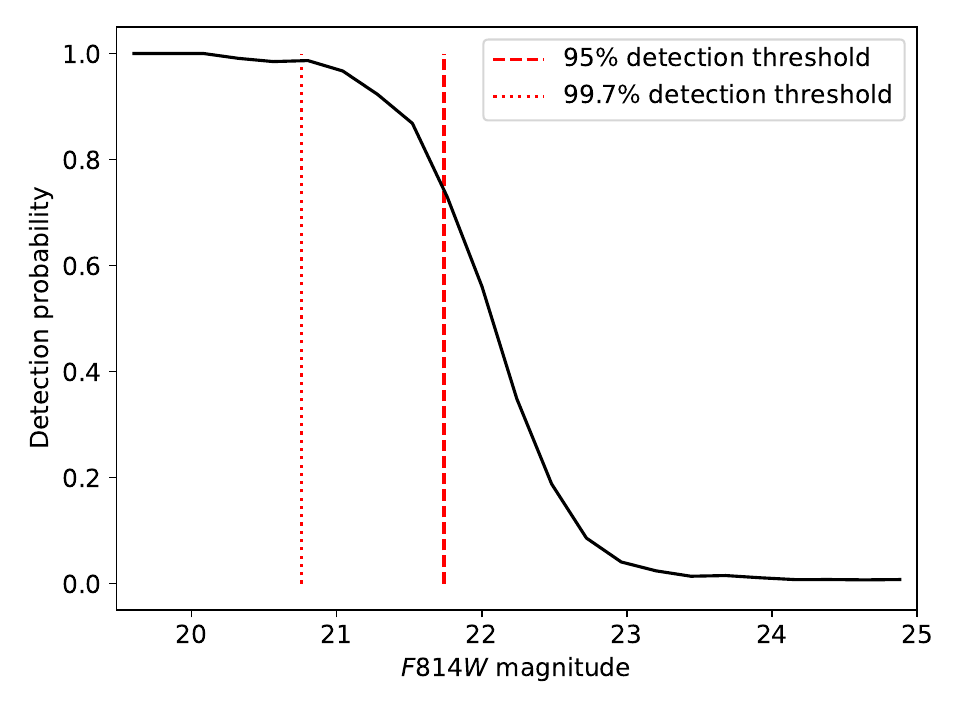}
    \caption{The dependence between the detection probability and the observed F814W magnitude with the detection thresholds overplotted. Results are marginalized 
    over the distribution of separations.}
    \label{fig:F814W_detect}
\end{figure}

Using the aforementioned simulations, we established that $\Delta m_t  = 2.17$ mag, which is translated to the limiting magnitude $m_t = 21.74$ mag.
Hence, $95\%$ of putative hosts brighter than $21.74$ mag should be detected.
A stricter limit is also calculated at the $99.7\%$ detection ratio ($3\sigma$), which corresponds to $\Delta m_t = 0.94$ mag and $m_t = 20.76$ mag.
The dependence between the observable luminosity of the star and the detection probability is presented in Figure \ref{fig:F814W_detect}.

\subsection{What percent of putative host-stars can we detect?}
In the end, we decided to simulate an artificial population of stars that would represent the potential population of a putative host stars.
The masses of the stars were sampled from the Chabrier mass function \citep{chabrier_galactic_2003}, 
while proper motions and distances were obtained using the Galactic model presented in \citet{batista_moa-2009-blg-387lb_2011}.
Then, the mass - $I$-band and the mass - $V$-band luminosity relationships presented in \citet{pecaut_intrinsic_2013} 
were used to obtain observable brightness in the F814W
filter with the transformation from \citet{holtzman_photometric_1995}. We
assume that the extinction is proportional to the total integrated density
of the gas in the line of sight.
The probability is assigned to each star based on the geocentric relative lens-source proper motion value $\mu_\textrm{rel}$ 
\begin{equation}
    P(\mu_g) = \frac{1}{\sqrt{2\pi \sigma^2}}\exp{-\frac{(\mu_\textrm{rel} - \mu)^2}{2\sigma^2}},
\end{equation}
where $\mu$ and $\sigma$ represent the mean and standard deviation of geocentric proper motion obtained from the microlensing curve and presented in Table \ref{tab:par_2}.

We decided to include an additional term in the probability that is related to the mass of the host star $P \propto M^n$. Several studies have found 
that the likelihood that a given star hosts a planetary system is a function of the mass of the star (see \citet{mulders_exoplanet_2018} for a review on the host star properties).
We decided to use the results presented in \citet{johnson_giant_2010}, where a nearly linear dependency on the host-star mass was found ($n = 1$).
We also decided to include in the analysis a population of host stars with no probability
dependence on the mass ($n= 0$) following \citet{mroz_free-floating_2024}.

\begin{figure}
    \includegraphics[width = 0.5\textwidth]{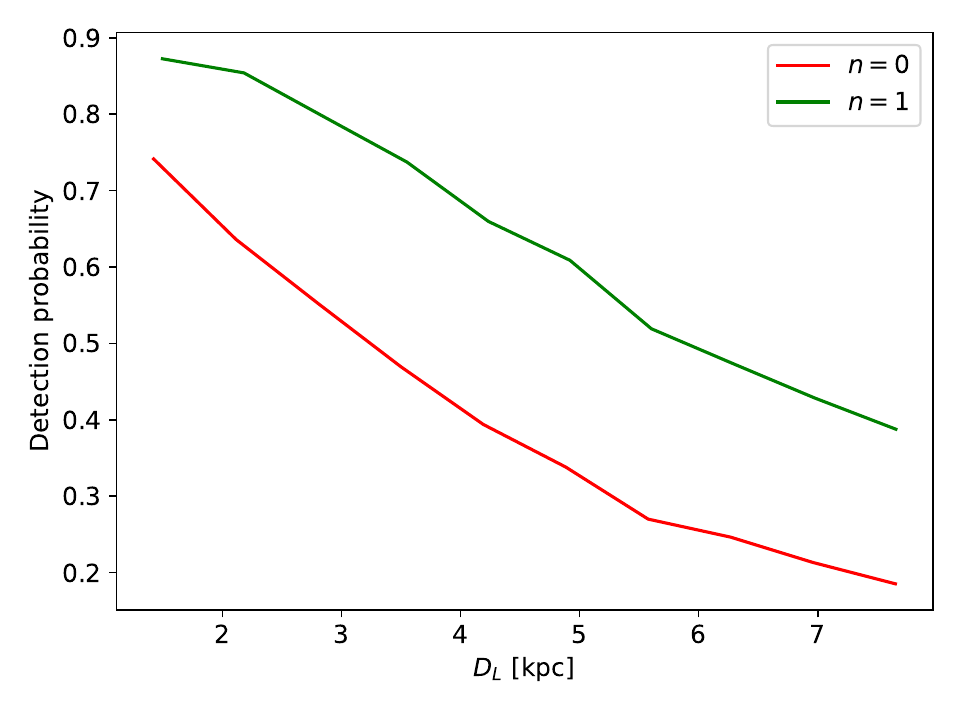}
    \caption{Dependence between the detection probability and the distance to the lens $D_L$.
    Two populations are considered: no probability dependence on mass ($n=0$) and $P \propto M$ ($n=1$).
    Results are marginalized over the distribution of magnitudes and separations.}
    \label{fig:d_l_prob}
\end{figure}

We took $10^7$ samples in total from the Galactic model. Subsequently, we selected $5\cdot 10^4$ with the Sampling, Importance, Resampling procedure according to the obtained probabilities. Then, we performed detectability simulations
for resampled objects according to the previously outlined procedure. This allowed us to model the detectability of putative hosts with proper priors on the separations and magnitudes.
It was determined that for $n=0$ only $25\%$ of stars should be detected. For $n=1$ this number is increased to around $48\%$. 
Hence, predicted detection ratios are comparable to values presented in \citet{mroz_free-floating_2024}. Moreover, the detection probability dependence on the distance to the lens 
was determined and is presented in Figure \ref{fig:d_l_prob}.

\begin{figure}
    \includegraphics[width = 0.5\textwidth]{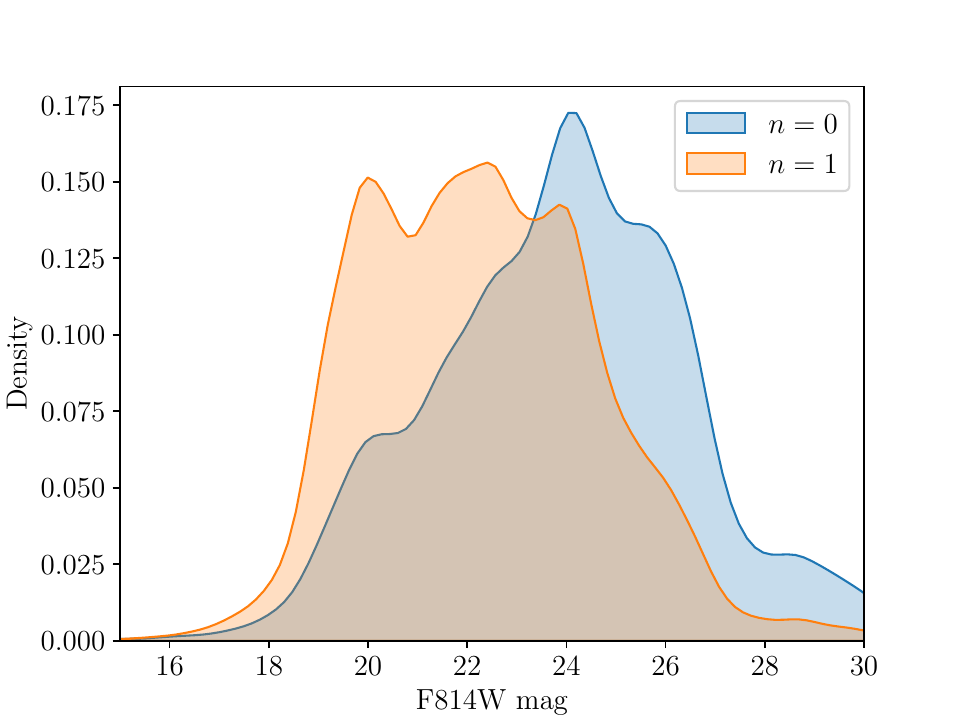}
    \caption{Probability density of F814W host stars magnitudes.
    Two populations are considered: $n=0$ and $n=1$.}
    \label{fig:hist}
\end{figure}

Despite a much longer baseline, no strict limits can be imposed. First of all, optical observations (even those in the $I$ band)
cannot detect most main sequence stars with such a short exposure time. To understand the limitations associated with the detections, let us consider 
the simulated distribution of F814W magnitudes presented in Figure \ref{fig:hist}.
Based on the HST images we conclude that stars fainter than $24$ mag are not detected on images in F814W filter.
We assume that this particular value is a limiting magnitude (fainter stars are impossible to detect no matter whether they are crowded or not).
This assumption allows us to estimate the fraction of putative host stars which should be detected without any other obscuring 
factors (like crowding).
Although in the $n=1$ case, we would be able to recover a significant part of the population (around $76\%$), for the $n=0$ case we would be able to recover only $50\%$ of those.
Hence, such a shallow image is not able to provide any conclusive answer with regard to the event's nature. In general, near-infrared observations with a telescope like JWST 
would be able to obtain much stricter limits than the presented WFPC2 analysis.

\section{Conclusions}
As proved in the previous section, it is highly unlikely that the source star is contaminated with additional blending from the putative lens-host
(even if, the total change in flux is negligible).
Hence, in the final analysis, we decided to use the prior on lens's $I$ magnitude (second column in Table \ref{tab:par_2}). The final mass of the lens can be computed with 
\begin{equation}
 M = \frac{\theta_\textrm{E}^2}{\kappa \pi_{\textrm{rel}}},
\end{equation} 
where $\kappa = 8.144\textrm{ mas}\,M_\odot^{-1}$.
Under the assumption that the lens is located in the Galactic disk (e.g., $\pi_{\textrm{rel}} = 0.1$ mas), one obtains $M = 9.3 M_{\oplus}$.
If the lens is located in the Galactic bulge (e.g., $\pi_{\textrm{rel}} = 0.016$ mas), one obtains $M = 58 M_\oplus$ ($0.6$ Saturn's mass).

OGLE-2023-BLG-0524 presents an interesting case, for which archival photometric measurements allowed us to obtain directly the color of the source star.
If no such observations had been made, the color would have been estimated from the $V$-band measurements during the peak.
However, due to the very short effective timescale of the event, $V$-band measurements are very limited (only one measurement from the KMTC site).
Despite the fact that the OGLE-2023-BLG-0524 event has a longer Einstein timescale than other FFP-candidate
events observed up to this day, the source-star crossing time $t_*$ is one of the shortest that have been measured.
In this publication we navigated the issue with the help of HST data, however, many similar microlensing events will lack color estimation.
This may pose a significant problem in the future as the
$\theta_\textrm{E}$ determination is heavily dependent on the source's color \citep{mroz_free-floating_2020}.

Hopefully, next-generation telescopes like the Extremely Large Telescope (ELT) would allow to mitigate such issues.
Let us consider an event similar to OGLE-2023-BLG-0524.
Assuming $\textrm{FWHM}=14.2 \textrm{ mas }\frac{\lambda}{2.2\mu \textrm{m }}/\frac{D}{39\textrm{ m }} \approx 14$ 
mas on the MICADO instrument ($K$ band, \citealt{MICADO}), three years would suffice for the source-lens system to separate to the extent of the FWHM width.
At the same time, a $39$ m mirror would allow one to detect much fainter hosts and enable the determination of the source's color.

The detectability simulations allow us to impose some constraints on the putative host star population. Despite a nearly $25$-year baseline, we are able to reject only 
around $25\%$ of potential stars. Hence, these archival HST observations turned out to have insufficient depth to conduct decisive detectability simulations.
We are able to constrain the maximum luminosity of the putative host star to around $21.74$ mag in the $I$ band at $95\%$ confidence interval.
However, our work points out that such archival observations create a viable way of veryfing the Free-floating origin of microlensing events.
We hope that telescopes like JWST, Euclid, or Nancy Roman Space Telescope 
will provide better opportunities in the future for such inquiries.

At the end, we estimated the number of microlensing events with such archival HST data. We used OGLE IV data presented in \citet{mroz_microlensing_2019},
obtaining the event rate density $\Gamma_{\textrm{deg}^2} = 113.9$ yr$^{-1}$deg$^{-2}$ for the BLG511 OGLE-IV field.
If we use all $3$ WFPC2 detectors ($80''\times 80''$ each), we expect around $0.17$ events per year.
Hence, many events should be present on the archival HST images, similar to the one used in this study. We conducted a search for events among the OGLE IV database
that could have been covered by this particular HST dataset. One event (OGLE-2017-BLG-0960) is located just a few pixels away from the edge of the detector. Nevertheless,
such searches for microlensing objects across archival HST datasets give a possibility to get a better view of the microlensing system, as it was many years before the event.

\begin{acknowledgements}
    We would like to thank Tomasz Bulik for the discussion.  
    This research was funded in part by the National Science Centre, Poland, grant OPUS 2021/41/B/ST9/00252 awarded to P.M.
    S.T. was supported in part by NASA under award number 80GSFC21M0002.
    This research is based on observations made with the NASA/ESA Hubble Space Telescope
    obtained from the Space Telescope Science Institute, which is operated by the Association of Universities for Research in Astronomy, Inc., under NASA contract NAS 5–26555.

    This work presents results from the European Space Agency (ESA) space mission Gaia.
    Gaia data are being processed by the Gaia Data Processing and Analysis Consortium (DPAC).
    Funding for the DPAC is provided by national institutions, in particular the institutions participating in the Gaia MultiLateral Agreement (MLA).
    The Gaia mission website is https://www.cosmos.esa.int/gaia. The Gaia archive website is https://archives.esac.esa.int/gaia.
    
    This research has made use of the KMTNet system
    operated by the Korea Astronomy and Space Science Institute
    (KASI) at three host sites of CTIO in Chile, SAAO in South
    Africa, and SSO in Australia. Data transfer from the host site to
    KASI was supported by the Korea Research Environment
    Open NETwork (KREONET). This research was supported by KASI
    under the R\&D program (project No. 2025-1-830-05) supervised
    by the Ministry of Science and ICT.

    % Individual acknowledgements
    W.Zang, H.Y., S.M., R.K., J.Z., and W.Zhu acknowledge support by the National Natural Science Foundation of China (Grant No. 12133005). 

    W.Zang acknowledges the support from the Harvard-Smithsonian Center for Astrophysics through the CfA Fellowship. 

    J.C.Y. and I.-G.S. acknowledge support from U.S. NSF Grant No. AST-2108414. 

    Work by C.H. was supported by the grants of National Research Foundation of Korea (2019R1A2C2085965 and 2020R1A4A2002885). 

    % For 2023 Events:
    J.C.Y. acknowledges support from a Scholarly Studies grant from the Smithsonian Institution.

\end{acknowledgements} 

\bibliographystyle{aa}
\bibliography{FFP}

\end{document}